\def\um{\mu}
\documentclass[aps,prb,twocolumn,superscriptaddress,amsmath,amssymb,floatfix]{revtex4-2}
\usepackage{graphicx}
\usepackage{color}
\usepackage{placeins}
\usepackage{xspace}
\usepackage{xcolor}
\usepackage{float}
\definecolor{red}{rgb}{0,0,0}
\usepackage{hyperref} 
\begin{document}
\title{Magnetic terahertz resonances above the N\'eel temperature in the frustrated kagome antiferromagnet averievite}
    \author{Tobias Biesner}
    \affiliation{1.~Physikalisches Institut, Universit\"at Stuttgart, Pfaffenwaldring 57, 70550 Stuttgart, Germany}
     \author{Seulki Roh}\email[Corresponding author: ]{seulki.roh@pi1.uni-stuttgart.de}
     \affiliation{1.~Physikalisches Institut, Universit\"at Stuttgart, Pfaffenwaldring 57, 70550 Stuttgart, Germany}
     \author{Andrej Pustogow}
     \affiliation{\textcolor{red}{Institute of Solid State Physics, TU Wien, 1040 Vienna, Austria}}

          \author{Hong Zheng}
     \affiliation{Materials Science Division, Argonne National Laboratory, Argonne, Illinois 60439, USA}
      \author{J. F. Mitchell}
     \affiliation{Materials Science Division, Argonne National Laboratory, Argonne, Illinois 60439, USA}
     \author{Martin Dressel}
    \affiliation{1.~Physikalisches Institut, Universit\"at Stuttgart, Pfaffenwaldring 57, 70550 Stuttgart, Germany}

    \date{\today}

\begin{abstract}
Time-domain magneto-THz spectroscopy is utilized to study the frustrated magnet averievite Cu$_{5-x}$Zn$_x$\-V$_2$O$_{10}$(CsCl). Pronounced THz resonances are observed in unsubstituted samples ($x=0$) when cooling below the onset of short-range magnetic correlations. The influence of external magnetic effects confirms the magnetic origin of these resonances. Increasing Zn substitution suppresses the resonances, as frustration effects dominate, reflecting the non-magnetic phases for $x> 0.25$ compounds. 
\textcolor{red}{The temperature evolution of the THz spectra is complemented with electron spin resonance spectroscopy. This comparison allows a
direct probe of the different contributions from magnetic order, frustration, and structural properties in the phase diagram of averievite.} 
Our results illustrate the effect of magnetic interactions in THz spectra of frustrated magnets.
\end{abstract}
\maketitle

\section{Introduction}
The fascination for the elusive quantum-spin-liquid state did not fade a bit albeit fifty years of research made significant contributions to its understanding. In quantum spin liquids (QSLs) no long-range magnetic order develops despite the presence of strong magnetic interactions; besides disorder and quantum effects, the main reason comes from geometrical frustration induced by the crystallographic lattice \cite{Balents2010,Broholm20,Inosov18,Mendels16,Hermanns18}.
For that reason, tuning frustrated magnetic systems provides a versatile toolbox for exploring the exotic ground states in the vicinity of
a QSL phase \cite{Knolle2019,Balents2010}.

Although extensive experimental \cite{Norman16,Little2017,Wu2018,Shi2018,Ponomaryov2020} and theoretical work \cite{Ng2007,Huh2013,Potter2013,Jeschke2013,Iqbal13,Ma2016,Bolens2018} has been performed in this field, \textcolor{red}{identifying fingerprints of QSL states \cite{Wen2019}, a direct experimental detection (smoking gun) remains elusive.} To that end, if we systematically tune the magnetic phase from a long-range order to a highly frustrated non-magnetic state, we expect useful insights on exotic magnetism \cite{Biesner2019}.

\textcolor{red}{Over the recent years, a plethora of candidate systems with different frustrated lattice geometries and magnetic interactions was successfully synthesized \cite{Balents2010}. Among them honeycomb materials (stripe/ zigzag antiferromagnets) are close to the Kitaev spin liquid, for instance $\alpha$-RuCl$_3$ \cite{Hermanns18}. Furthermore, the kagome lattice is suggested to harbor rich magnetic phases: U(1), or $Z_2$ spin liquids, $Q=0$, or $\sqrt{3} \times \sqrt{3}$ states \cite{Mendels16, Norman16}. Especially herbertsmithite [ZnCu$_3$(OH)$_6$Cl$_2$] \cite{Shores05} has been highlighted as a QSL candidate \cite{Norman16}. Recently, the interest in distorted kagome systems has grown as well: Y-kapellasite [Y$_3$Cu$_9$(OH)$_{19}$Cl$_8$, $Q=(1/3,1/3)$ magnetic structure] \cite{Puphal17, Hering2022}, volborthite [Cu$_3$V$_2$O$_7$(OH)$_2$$\cdot$$2$H$_2$O, spin trimer] \cite{Kohama19}, or Rb$_2$Cu$_3$SnF$_{12}$ (pinwheel valence bond solid) \cite{Matan2010}. These systems could develop exotic (noncollinear) magnetic phases with an interplay of lattice geometry and quantum and thermal fluctuations and offer great tunability in the proximity of QSLs \cite{Hering2022, Norman2020}.}

\begin{figure}[!t]
\centering
\includegraphics[width=0.94\columnwidth]{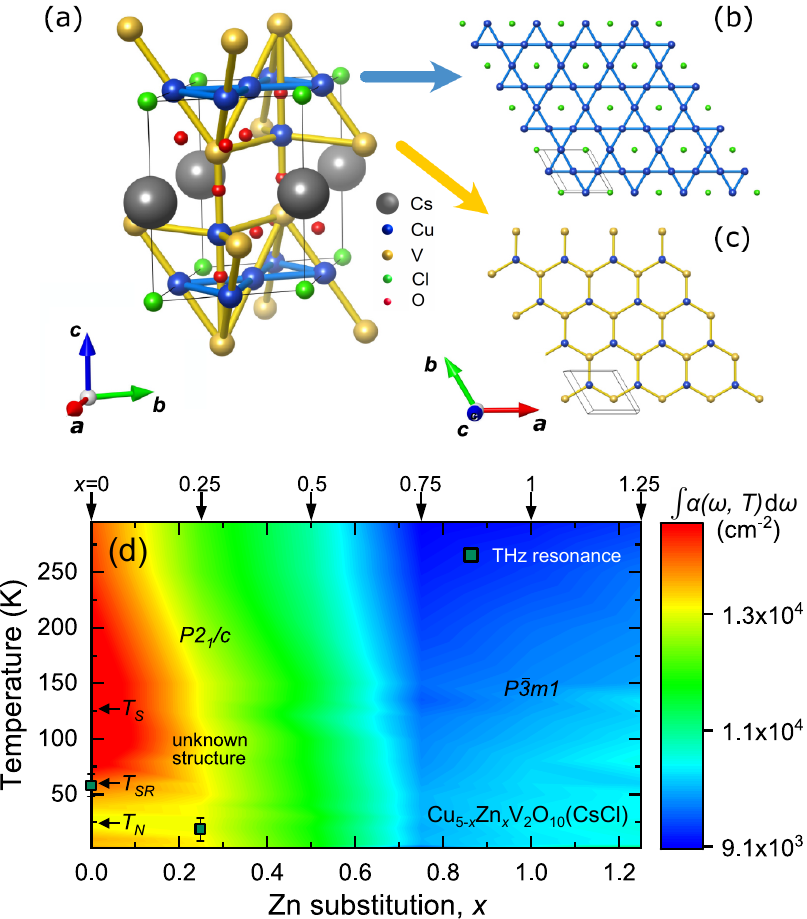}
\caption{\label{Fig:structure} Crystal structure of the averievite Cu$_{5-x}$Zn$_x$\-V$_2$O$_{10}$(CsCl). The $P\bar{3}m$1 structure is
found for $x=0$ only for $T> 310$~K while for $x=1$ in the entire temperature range.
(a)~A single unit cell representation. Two different in-plane ($ab$) sheets are stacked along the $c$ direction. One is the kagome structure (b) composed of Cu atoms and the other is a honeycomb lattice (c) consisting of V and Cu. Overall, averievite crystallizes in alternating kagome and two honeycomb lattices. (d) {\color{red}Structural and magnetic properties of averievite. $T_S=127$~K, structural transition; $T_{\rm SR}\approx 60$~K, onset of short-range correlations; $T_{N}=24$~K, long-range order. False-color plot is based on the integrated absorption coefficient $\int\alpha (\omega, T)\,\mathrm{d}\omega$ obtained for different Zn concentrations (indicated by vertical arrows at the top). Dark green points: Onset of magnetic THz resonances.}}
\end{figure}

Averievite, Cu$_{5}$V$_2$O$_{10}$(CsCl), constitutes a very particular case \cite{Vergasuva98,Starova97,Botana18,Volkova2018}
because a frustrated Cu$_3$O$_2$Cl kagome layer is sandwiched between CuVO$_3$ honeycomb sublattices [see Fig.~\ref{Fig:structure}(a-c)].
The coexistence of a second distinct layer offers two distinguishable copper sites with a square planar and a trigonal bipyramidal coordination for the kagome and honeycomb layers, respectively. Above $T = 310$~K, the trigonal $P\bar{3}m$1 structure features an isotropic kagome network. However, upon cooling, the system enters a monoclinic $P2_1/c$ spacegroup, slightly distorting the kagome geometry. The second transition occurs, below  $T_S = 127$ K, to an unresolved structure \cite{Botana18}. Susceptibility measurements reveal strong antiferromagnetic couplings with a Curie-Weiss temperature of $\Theta_{CW} = 185$ K.

However, the compound freezes into long-range order only below $T_N = 24$ K, suggesting a moderate degree of frustration and the onset of short-range magnetic correlations well above $T_N$ [\textcolor{red}{see} Fig.~\ref{Fig:structure}(d)]. \textcolor{red}{The magnetic entropy assigned with this transition ($\Delta S_{mag} = 1.1$ J/mol K) represents only 3.8 \% of the expected $R$ ln(2), pointing to remaining spin fluctuations even in the ordered phase. Moreover, susceptibility measurements do not show any dependence on the cooling protocol (field-cooled/ zero-field-cooled data), excluding the possibility of a spin-glass transition \cite{Botana18}. Theoretical investigations suggest a herringbone valence band solid \cite{Dey20,Norman2020}.} Magnetic nearest-neighbor coupling was calculated to be the dominant magnetic exchange of $J_1  = 227.8$ K within the kagome layer \textcolor{red}{(in-plane)} and one along the out-of-plane, between kagome and honeycomb lattice, as $J_2=51.7$ K \cite{Dey20}, \textcolor{red}{consistent with the experimentally determined Curie-Weiss temperature \cite{Botana18}.} Further magnetic coupling between the two kagome layers is sufficiently suppressed due to intercalated CsO$_2$ sites.

An additional degree of freedom is provided by chemical substitution; to that end different strategies were explored theoretically and experimentally \cite{Botana18,Winiarski19,Dey20}. Among them, exchanging Cu by non-magnetic Zn in Cu$_{5-x}$Zn$_x$V$_2$O$_{10}$(CsCl)
was found to replace the copper ions in the honeycomb layers. This substitution reduces the magnetic interlayer coupling preserving the highly frustrated $S=1/2$ kagome physics with no indication of long-range order \textcolor{red}{or spin-glass behavior} down to $T=2$~K. Furthermore, in Cu$_{4}$Zn$_1$V$_2$O$_{10}$(CsCl) ($x=1$) the monoclinic distortion upon cooling is absent, such that the compound remains in the more symmetric $P\bar{3}m$1 structure with an undistorted kagome lattice down to the lowest measured temperature [cf. Fig.~\ref{Fig:structure}(d)]. \textcolor{red}{The specific heat shows a gapless behavior with a $C_p/T\sim T^2$ temperature dependence below $T=10$~K \cite{Botana18}. Such a behavior might be expected for a U(1) QSL.} This not only promotes Zn-substituted averievite as a QSL candidate, but the entire averievite family can be taken as a platform for investigating the magnetic ground state under frustration.

In the present work, we explore the physics of Cu$_{5-x}$Zn$_x$V$_2$O$_{10}$(CsCl) by employing THz time-domain spectroscopy (THz-TDS). This method is a powerful tool to map out the electromagnetic response of a material in the THz-frequency range. For insulators, magnetic properties and phonon contributions are best visible at low frequencies as the electronic background becomes diminished \cite{Pan2014, Lin2015, Kim2015, Guo2020,Higuchi2011,Cheng14,Qiu2020}. Of particular interest here are magnetic resonances \cite{Kittel1951,Keffer1952} that are induced by the magnetic field of the THz light over a Zeeman torque. When probed by THz-TDS these excitations show up as a free induction decay with a characteristic time scale \cite{Kirilyuk10,Yamaguchi2010,Yamaguchi2013,Kubacka1333,Nakajima2010,Hansteen2005,Lu2017,Nishitani2010}.
For $x=0$, Cu$_{5}$V$_2$O$_{10}$(CsCl), we observe multiple sharp resonance modes that develop with the onset of short-range magnetic interactions ($T_{\rm SR} \approx 60$~K) with pronounced extended-time oscillations, a fingerprint of magnetic THz resonances \cite{Yamaguchi2010,Yamaguchi2013}. Our magneto-THz investigation further clarifies the magnetic origin of these features. As frustration disturbs the ordered ground state in Zn-substituted averievite, magnetic resonances get suppressed, leaving a featureless response (other than the electronic/phononic background) reminiscent of quantum spin liquids \cite{Pilon13, Potter2013}.

\section{Experimental Methods}
Averievite powders of various substitution rates ($x= 0$, 0.25, 0.50, 0.75, 1, 1.25) were prepared and characterized as described previously \cite{Botana18}. The THz measurements were performed in transmission geometry. Pellets with a typical thickness of $100~\mu$m were pressed from fine powder. All specimens were measured with a time-domain THz spectrometer (TeraView TeraPulse 4000) attaching a homemade He-bath cryostat and a superconducting magnet (Oxford Instruments). The magneto-THz experiments were performed in Faraday geometry (see \cite{SM} for further information). The recorded time varying electric fields were further analyzed by fast Fourier transformation (FFT) to obtain the frequency-domain spectra. From that we calculated the absorption coefficient $\alpha(\omega)$ via the well-known Beer-Lambert law:
$\alpha(\omega) = -\ln\{Tr (\omega)\}/d$, where $Tr$ is the transmittance and $d$ the sample thickness \cite{tanner2019}. The integration of $\alpha (\omega)$ over  frequency, \textcolor{red}{$\int\alpha (\omega)\,\mathrm{d}\omega$}, thus is proportional to the optical spectral weight \cite{dressel2002}.

Furthermore, electron spin resonance (ESR) measurements were conducted via a continuous-wave X-band spectrometer (Bruker EMXplus) equipped with a He-flow cryostat (Oxford Instruments, ESR 900) at 9.8~GHz. The spin susceptibility $\chi_s^e$ is obtained by double integration of the derivative of the microwave power with respect to the magnetic field, d$P$/d$H$, typically recorded as a function of temperature $T$.

\begin{figure}
\centering
\includegraphics[width=1\columnwidth]{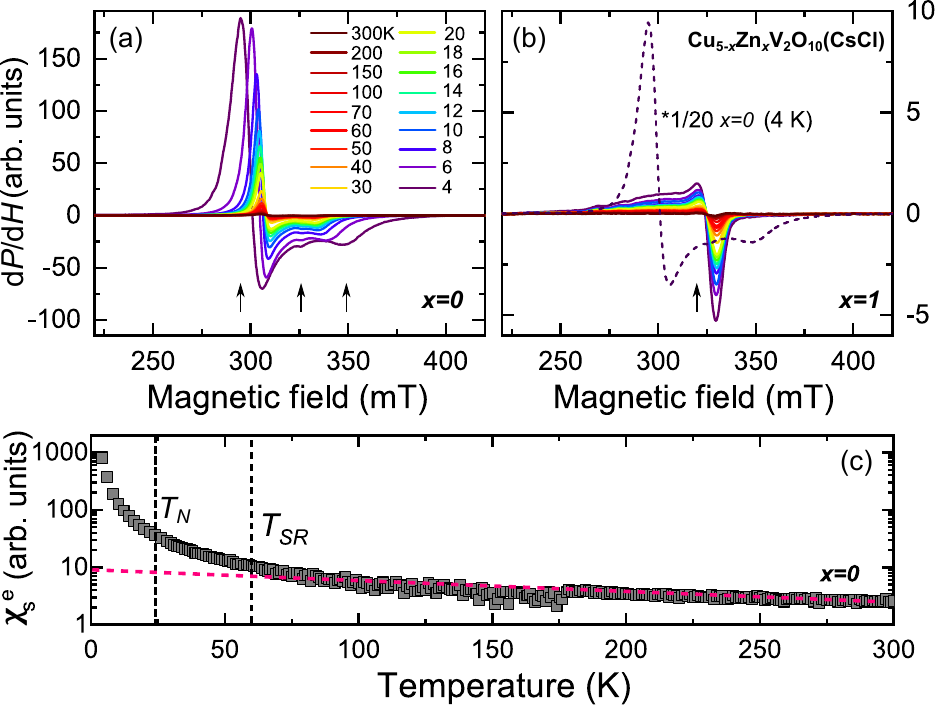}
\caption{\label{Fig:ESR} (a,b) X-band ESR spectra of averievite with $x=0$ and $x=1$ (derivative of the absorbed microwave power with respect to the magnetic field) recorded at different temperatures as indicated. Upon cooling below 60~K, several strong Lorentzian modes are observed for $x=0$. The spectral weight is significantly suppressed for the $x=1$ compound. Panel (c) shows the electron-spin susceptibility $\chi_s^e(T)$ for the $x=0$ compound obtained from ESR measurements. Vertical black dashed lines represent the $T_N=24$~K and $T_{\rm SR}\approx 60$~K. Colored dashed line represents the change of trend under the onset of short-range magnetic correlations.}
\end{figure}

\section{Results and Discussion}
To characterize the magnetism of averievite, temperature-dependent ESR measurements in the X-band frequency were performed for $x=0$ and $x=1$ powders. The results are presented in Fig.~\ref{Fig:ESR}(a,b).
In the $x=0$ sample, three distinct Lorentzian modes can be clearly identified at $T\approx 60$~K, and their intensity strongly increases upon cooling. This splitting indicates the different magnetic contributions from kagome and honeycomb copper sites. Substituting Cu by Zn ($x=1$) leads to an overall suppression of spectral weight, i.e., a considerable change of magnetism for Zn-substituted samples. This result is in accord with susceptibility measurements \cite{Botana18}, reporting a suppression of the magnetic order as the Zn concentration increases. Fig.~\ref{Fig:ESR}~(c) displays the temperature-dependent electron-spin susceptibility $\chi_s^e$ for the $x=0$ compound, extracted from the ESR spectra. When $T$ decreases, $\chi_s^e(T)$ starts to rise around 60~K, indicating the onset of short-range magnetic correlations\textcolor{red}{/ short-range order}. It further amplifies close to $T_N=24$~K as magnetic interactions stabilize. These results yield two characteristic temperatures for magnetic properties of $x=0$ averievite: the onset of short-range correlations, $T_{\rm SR}\approx60$~K, and long-range order, $T_N=24$~K; see Fig.~\ref{Fig:structure}(d).

\begin{figure}
\centering
\includegraphics[width=1\columnwidth]{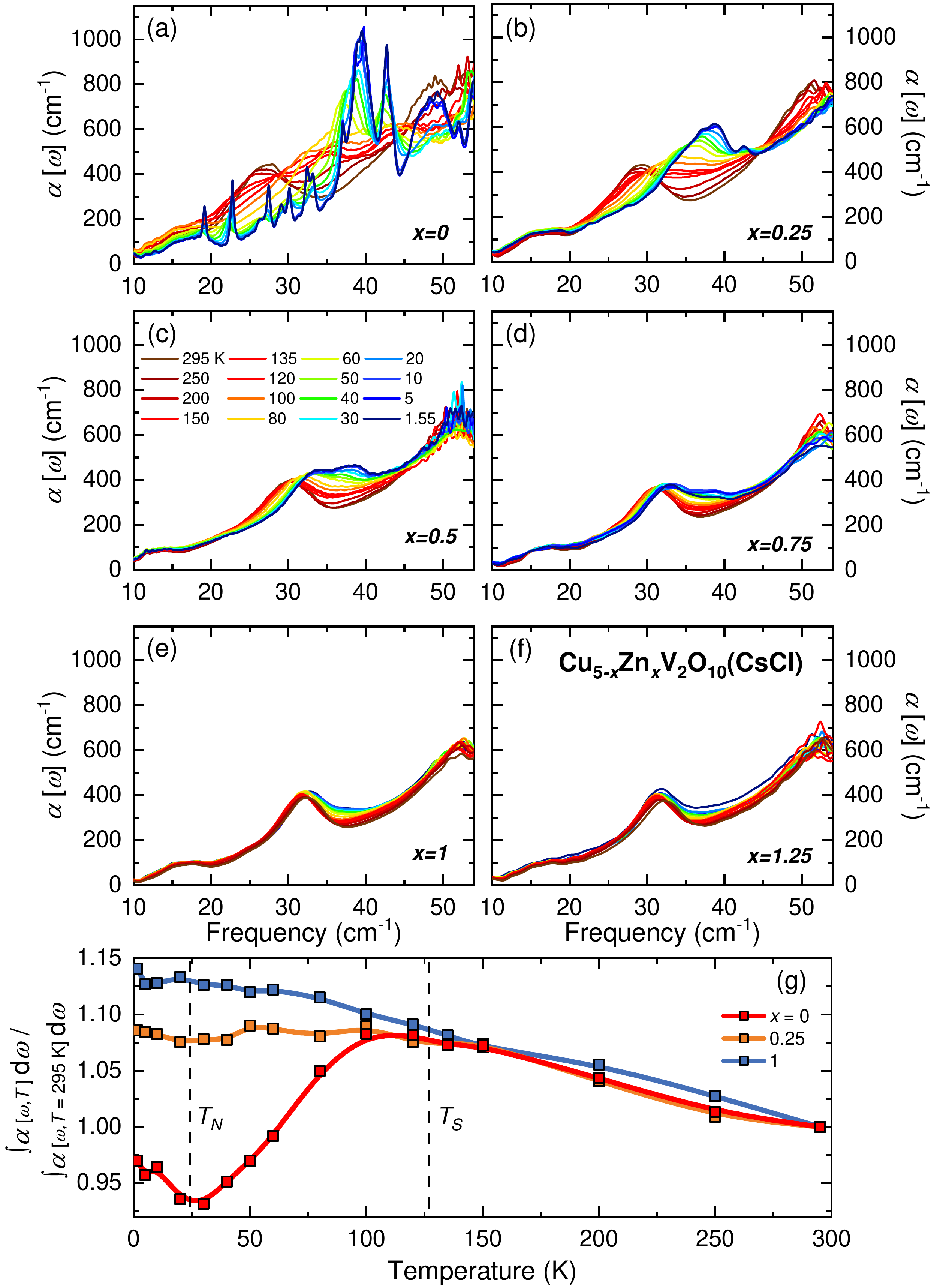}
\caption{\label{Fig:spectra}(a-f) Absorption coefficient, $\alpha (\omega)$ spectra of a series of Cu$_{5-x}$Zn$_x$V$_2$O$_{10}$(CsCl) samples with varied Zn contributions, obtained from THz-TDS. At room temperature, all compounds contain similar peaks at around 15, 28, and 50 cm$^{-1}$. Upon cooling through the structural transition, the lower concentrations $x<0.75$ show pronounced changes. For $x=0$ several sharp resonances develop below 60 K. (g) Frequency-integrated absorption coefficient (10 to 50 cm$^{-1}$) as a function of temperature for representative substitution, normalized by the 295 K data. Two vertical dashed lines represent $T_N$ and $T_S$, respectively.}
\end{figure}

The frequency-dependent absorption coefficient $\alpha (\omega)$, obtained by THz time-domain spectroscopy,
is presented in Fig.~\ref{Fig:spectra} for different Zn substitutions $x$. Overall, the spectra are characterized by an insulating response. At elevated temperatures, three maxima are present at around 15, 28, and 50~cm$^{-1}$ for the $x=0$ compound [Fig.~\ref{Fig:spectra}(a)]. These features most likely stem from low-energy phonon modes because at THz frequencies no electronic contributions are expected due to the highly insulating nature of averievite \cite{Botana18}. Furthermore, these phonon modes are affected by Zn substitution (see \cite{SM}). Here, the center frequencies are slightly higher compared to the unsubstituted compound. Especially, the 28~cm$^{-1}$ peak exhibits a slight blueshift, likely due to the combined effect of Zn substitution on the lattice and different space groups of the $x=0$ and $x=1$ samples. The blueshift is expected due to the reduced bonding length (increased bonding force) of Zn compared to Cu. \textcolor{red}{This lets us speculate that the phonons involve honeycomb and kagome sites (out-of-plane vibrations) or are associated with the distortion of the kagome lattice.}

Let us first focus on the $x=0$ spectra  displayed in Fig.~\ref{Fig:spectra}(a), where we can see that the two latter peaks exhibit a significant temperature dependence when cooling down to $T \approx 100$~K. While the 28~cm$^{-1}$ mode hardens, the 50~cm$^{-1}$ feature moves to lower frequencies resulting in a broad contribution centered at around 40~cm$^{-1}$. This broadening of the phonon modes upon cooling implies a lattice instability in the vicinity of the structural transition ($T_S = 127$~K). The broad phonon contribution sustains its frequency and its intensity continuously grows with cooling down to 1.55~K.

Drastic changes are observed in $\alpha(\omega)$ when the temperature is reduced below 60~K, where short-range magnetic correlations appear: multiple well pronounced and sharp peaks emerge distributed over the entire THz range down to 19~cm$^{-1}$. Given the spectral form and rather abrupt temperature evolution of these features, a pure phononic origin seems very unlikely. More details will be discussed later. Interestingly, samples with intermediate substitution [$x= 0.25$ and 0.5, presented in Fig.~\ref{Fig:spectra}(b,c)] undergo a comparable phonon evolution implying that the structural instability still resides in these samples, possibly exhibiting a similar structural phase transition (weaker but with the same trend as in $x=0$). The sharp resonances of the $x=0$ sample, however, are totally absent when Zn substitution exceeds $x=0.25$; cf.\ Fig.~\ref{Fig:spectra}(b-d). For high concentrations $x=1$ and $x=1.25$ [Fig.~\ref{Fig:spectra}(e,f)], only a negligible temperature evolution of the phonon modes is observed, suggesting that no structural transition occurs with cooling. This is in accordance with the synchrotron powder diffraction \cite{Botana18}. These trends can also be followed in the integrated absorption coefficient, \textcolor{red}{$\int\alpha (\omega, T)\,\mathrm{d}\omega/\int\alpha (\omega, T=295~{\rm K})\,\mathrm{d}\omega$}, displayed in Fig.~\ref{Fig:spectra}(g). Indeed, for the $x=0$ sample the temperature dependence clearly changes around $T_S = 127$~K.
Furthermore, the integrated absorption coefficient increases below $T_N=24$ K indicating a close connection to the sharp peaklike features and the magnetic degree of freedom observed in the unsubstituted sample. For higher substitution rates, the absorption curves are rather monotonic without any noticeable change. This also reflects the fact that Zn substitution suppresses the structural transition as well as magnetic order.

\begin{figure}
\centering
\includegraphics[width=1\columnwidth]{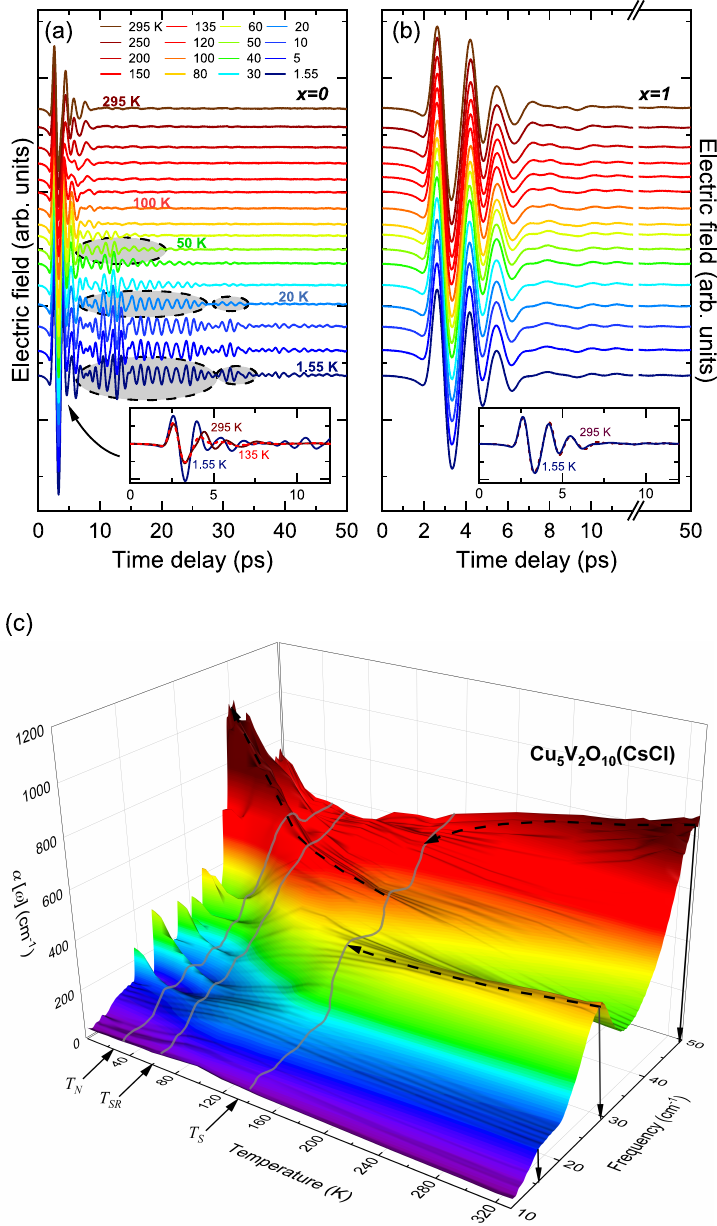}
\caption{\label{Fig:oscillations} Time-domain THz electric field transmitted through averievite pellet.
(a) and (b) represent the $x=0$ and $x=1$ samples of  Cu$_{5-x}$Zn$_x$V$_2$O$_{10}$(CsCl). The insets are the magnified views of the main echo signals. Below $60$~K pronounced extended-time oscillations develop for the $x=0$ sample [shaded area in (a)], well separated from the main pulse. For the $x=1$ sample these oscillations are absent. (c) False-color plot of the absorption coefficient $\alpha(\omega,T)$ spectra for the $x= 0$ sample. Three lines are representing the structural transition at $T_S=127$~K, onset of short-range magnetic correlations at $T_{\rm SR}\approx$ 60~K, and magnetic long-range order at $T_N=24$~K, respectively. The dashed lines trace the 28 and 50~cm$^{-1}$ modes.}
\end{figure}

In searching for the origin of the sharp low-temperature features of the $x= 0$ sample, we analyze the electric field of transmitted THz light in the time domain in more detail, Fig.~\ref{Fig:oscillations}(a).  At ambient temperature, the spectrum consists of the main pulse of the THz electric field, centered at around 3 ps. Decreasing the temperature below $T_{\rm SR}\approx60$~K, additional oscillation features  appear with longer periods (about 30 ps).
Cooling down further to $T=1.55$~K, these oscillations become more prominent. Especially below $T_N=24$~K, an additional group of oscillations can be identified between 30 to 35 ps, increasing the sharpness of the peaklike features in Fig.~\ref{Fig:spectra}(a).
The FFT analysis clearly shows that the phononic high-temperature response, i.e., the 15, 28, and 50~cm$^{-1}$ features, is rather confined to the main pulse.
However, the multiple sharp peaks in the low-temperature spectra correspond to the pronounced extended-time oscillation of the electric field. For the $x=1$ sample these oscillations are absent, Fig.~\ref{Fig:oscillations}(b), rendering a purely phonic response.

For the $x=0$ compound, a three-dimensional false-color plot of $\alpha(\omega,T)$ is presented in Fig.~\ref{Fig:oscillations}(c). Here, one can clearly separate the effects of the structural transition at $T_S = 127$~K, the spectral evolution in the vicinity of the short-range magnetic correlations, $T_{\rm SR}\approx60$~K, and the one related to long-range order below $T_N = 24$~K.
The structural transition leads to a redistribution of spectral weight from the two phonon modes at 28 and 50~cm$^{-1}$ to a new mode at 40~cm$^{-1}$.
At $T_{\rm SR}$ the sharp peaklike contributions appear and becomes stronger toward lower temperatures.
Some of the peaks (for instance, the one around 30~cm$^{-1}$) show additional sharpening when entering the long-range magnetically ordered phase below $T_N$.
In general, however, only monotonic changes can be observed at $T_N$.

It is notable that the sharp spectral peaks --~due to extended-time oscillations~-- emerge at the same temperature as magnetic short-range correlations. This fact is taken as strong evidence for the magnetic origin of these features.
Indeed, THz-driven spin precession (magnetic resonance) has recently been reported in several magnetically ordered systems \cite{Yamaguchi2010,Yamaguchi2013,Nakajima2010,Hansteen2005,Lu2017,Nishitani2010}. Here, the THz magnetic field adds a torque to the spins, resulting in a coherent precession of the magnetic moments.
The emitted electromagnetic wave can be detected via a time-gated detection scheme, which allows us to distinguish electronic and magnetic contributions by different timescales.
Although the total magnetization vanishes for antiferromagnetic systems, the magnetic resonance is still expected to be driven by the staggered magnetic component, i.e., the N\'{e}el order $\boldsymbol{n}$ \cite{Cheng14}. The electromagnetic field emitted during the free induction decay has a distinguishable oscillation period regarding the decaying time of the precession, which is on the order of several tens of ps, whereas the electronic response of the material is confined to shorter times of the incident THz light (a few ps) \cite{Kirilyuk10}.
The time-domain spectra of the $x=0$ compound [Fig.~\ref{Fig:oscillations}(a)] show such a pronounced extended-time delay signal.
The close temperature relation with the magnetic characteristic temperatures ($T_{\rm SR}$ and $T_N$) supports this idea.

At this point, we cannot fully exclude a possible magnetoelastic coupling mechanism from be involved. Such a scenario is used to explain splitting of the phonons under the emergence of magnetic order in other systems \cite{Sushkov2005,Kant2009}. \textcolor{red} {In the present case, a lattice vibration might modulate the spin configuration, i.e., a transient change of the distortion of the kagome lattice/ coupling of kagome and honeycomb copper sites. Especially, the contributions of out-of-plane (kagome-honeycomb) modes might favor the latter.} However, in distinction from a magnetic resonance, these splittings are expected to be, \textcolor{red}{most likely,} continuously developing from the original phonon modes, which is not the case for averievite [cf. Fig.\ref{Fig:oscillations} (c)]

\textcolor{red} {Another possibility is Brillouin zone folding with the development of a magnetic supercell at $T_N$ \cite{Sekine1990}. This could as well result in new THz features as the phonon wavelength increases, i.e., under unit cell doubling. Similar observations have been made, for instance, at the spin-Peierls transition \cite{Room2004} or charge order \cite{Samnakay2015}. However, our structural/ magnetic characterizations do not give any indications of such cases. Raman or infrared studies, providing a symmetry aspect, could give further insight about this relation of structural and magnetic degrees of freedom.} 

Furthermore, results of our ESR measurements on the $x=0$ compound match well with the magnetic temperature scale; the absorption starts to appear at $T_{\rm SR}$ and amplifies further at $T_N$. In addition, we observe multiple Lorentzian contributions in ESR result. In this case, the system might contain distinct magnetic contributions, i.e., multiple magnetic channels (different Cu sites). Interestingly, in our THz measurements we observe multiple resonances as well, inferring the superposition of several oscillations in the time-domain signal.

\begin{figure}
\centering
\includegraphics[width=1\columnwidth]{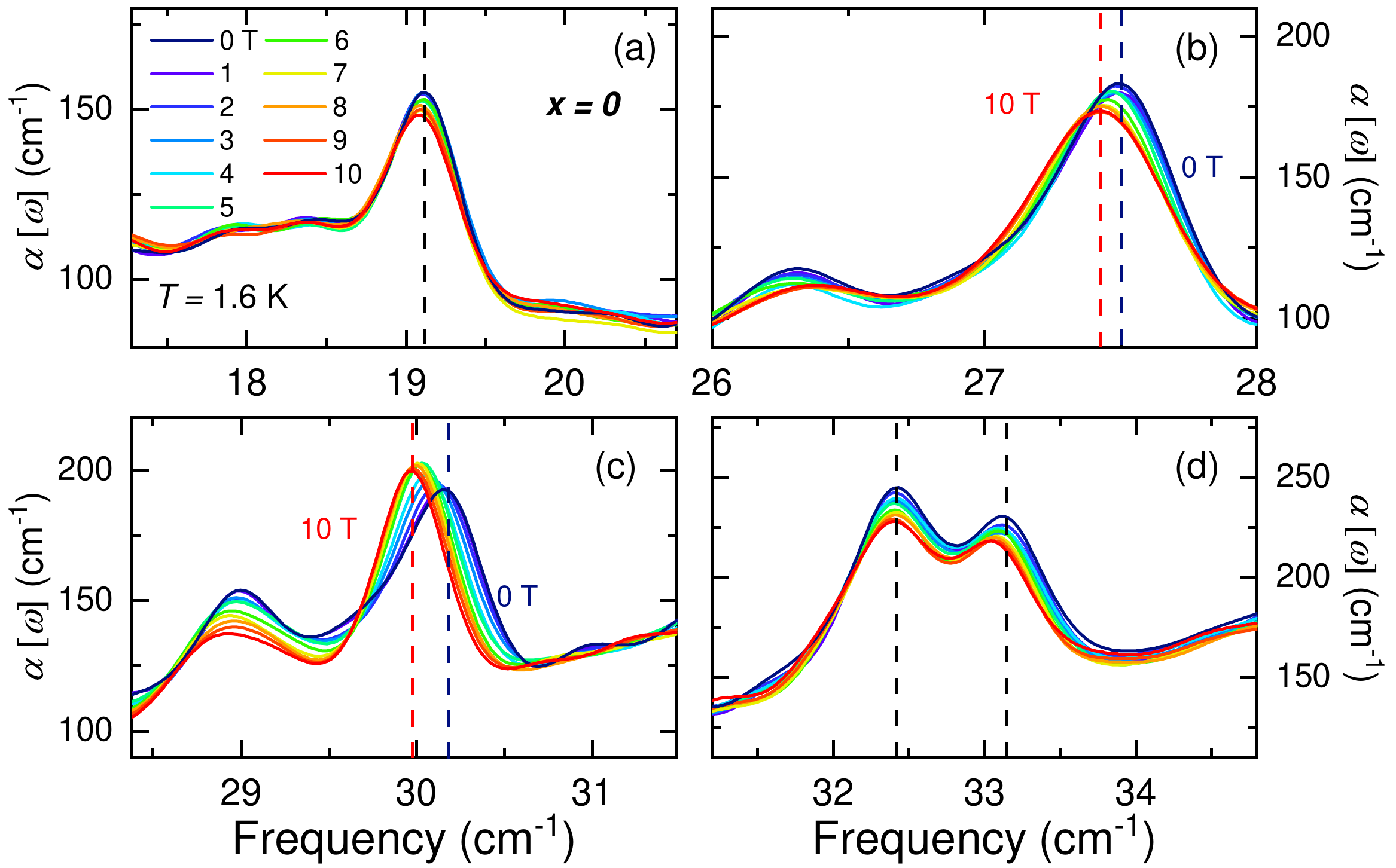}
\caption{\label{Fig:magnetooptics} (a-d) Evolution of the frequency-dependent absorption coefficient $\alpha(\omega)$  under external magnetic field for Cu$_{5}$V$_2$O$_{10}$(CsCl)  at $T=1.6$~K. The black dashed lines indicate those modes which do not shift in field, but only modify their intensity.
The ones with both intensity and frequency shifts are marked at the center frequency at zero field (blue) and 10 T (red) with dashed lines [panel (b,c)].}
\end{figure}

To further confirm the origin of the low-temperature resonances in unsubstituted averievite, we carried out mag\-neto-THz measurements; the results are presented in Fig.~\ref{Fig:magnetooptics}. Some of the sharp features clearly shift to lower energies with magnetic field [see panels (b,c)].
In the Supplemental Material we demonstrate that this shift can already be concluded from modifications in the time-domain oscillations \cite{SM}.
Further investigations are necessary to firmly conclude on the final reason for the redshift under external magnetic field.
There are different possibilities such as enhanced magnon scattering under magnetic field \cite{Mukai2016, Suhl1957} or a slight adjustment of the internal staggered field along the external static field and thus a change in the resonance frequency and intensity.
Importantly, our magneto-THz measurements confirm the magnetic origin of the sharp resonances at 27.5 and 30 cm$^{-1}$.
Furthermore, the simultaneous appearance  of multiple low-energy resonances far below the structural transition suggests a common magnetic origin of all resonances.

Based on our comprehensive THz study we now gain some understanding of the structural and magnetic properties of averievite \textcolor{red}{from a spectroscopic view}. \textcolor{red}{Especially, with the false-color plot of $\int\alpha (\omega, T)\,\mathrm{d}\omega$, Fig.~\ref{Fig:structure}(d), we find a good agreement to previous static magnetometries and structural characterization \cite{Botana18}.}
The transition at $T_S=127$~K is indeed detected in the phonon spectra, where features merge and new phonons appear for the $x = 0$ sample.
For intermediate substitutions ($x= 0.25$ and 0.5), a similar phonon evolution can be revealed, reminiscent of the structural phase transition in the $x = 0$ specimen.
However, the phonon modes in Cu$_{4}$ZnV$_2$O$_{10}$(CsCl) ($x=1$) do not show any significant temperature dependence, providing evidence that the structural transition is absent.
As far as the magnetic aspects are concerned, we detect well pronounced magnetic resonances in the THz range for the $x=0$ averievite. It is important to point out that these features concomitantly appear with the onset of short-range magnetic correlations, even above the temperature $T_N$ before long-range order is established. This indicates the existence of strong magnetic interactions, i.e., paramagnons \cite{LeTacon2011,Qin2017}.
The magnetic features vanish with substitution as $x>0.25$. Increasing the Zn content further completely suppresses the magnetic order.
The THz response of $x=1$ averievite is not susceptible to an external magnetic field (see \cite{SM}),
in conjunction with the expectation of QSLs \cite{Pilon13,Potter2013}.

Previous DFT calculations suggest that Zn substitution specifically replaces the Cu ions within the honeycomb lattice \cite{Botana18}. In return, this substitution reduces the magnetic coupling between the adjacent kagome and honeycomb lattice leading to magnetically isolated kagome layers. Here, the magnetic fluctuations become stronger as the highly frustrated kagome layers decouple from each other, resulting in a suppression of magnetic order. The structural transitions are suppressed as well. Thus, the $x=1$ system remains in the $P\bar{3}m1$ space group down to the lowest measured temperatures without magnetic ordering. These results are well corroborated by our THz measurements, revealing the entanglement of structural an magnetic factors in the phase diagram of averievite.

\section{Conclusion}
In summary, we performed extensive magneto-THz measurements on averievite Cu$_{5-x}$Zn$_x$V$_2$O$_{10}$(CsCl). We detected magnetic resonances for the unsubstituted compound due to the spin precession induced by THz light. \textcolor{red}{We find a rather wide temperature range of magnetic short-range correlations $T_{SR}\approx 60$ K above long-range order $T_N = 24$ K. In fact, the THz resonances are coupled to the short-range order, similarly to paramagnons.} For samples substituted with a higher Zn concentration, frustration effects of the kagome lattice take over, leading to a suppression of the magnetic resonances; \textcolor{red}{they vanish when exceeding $x=0.25$ substitution}. \textcolor{red}{Corroborating the magneto-THz results our ESR investigation shows a suppressed spectral weight as well.} With the comprehensive information provided by THz spectra obtained in time domain we can \textcolor{red}{identify magnetic and structural transitions of averievite Cu$_{5-x}$Zn$_x$V$_2$O$_{10}$(CsCl)}. This study sheds light on the way magnetic interactions affect THz spectra \textcolor{red}{and magnetization dynamics} of frustrated magnets. We clear the way for directly probing the different contributions from magnetic order, frustration, and structural properties. Moreover, tracing the magnetic resonances and \textcolor{red}{their dynamics} gives an augmented, spectroscopic aspect of frustrated magnetism in the vicinity of QSLs \textcolor{red}{proving the efficacy of THz time-domain spectroscopy in this field.}

\section*{Acknowledgments}
We thank Antia S. Botana for fruitful discussion and Gabriele Untereiner for continuous \textcolor{red}{technical support}.
This project was supported by the Deutsche Forschungsgemeinschaft (DFG). Work in the Materials Science Division of Argonne National Laboratory (sample preparation, structural and magnetic characterization) was supported by the U.S. Department of Energy, Office of Science, Basic Energy Sciences, Materials Science and Engineering Division.
\FloatBarrier
\bibliographystyle{apsrev4-2}
\bibliography{literature}

\pagebreak

\setcounter{figure}{0}
\renewcommand{\thefigure}{S\arabic{figure}}
\renewcommand{\theHfigure}{Supplement.\thefigure}
\pagenumbering{gobble}
\widetext

\begin{center}
\textbf{\large Supplemental Materials for "Magnetic terahertz resonances above the N\'eel temperature in the frustrated kagome antiferromagnet averievite"}
\end{center}

\section{Zinc substitution}
\begin{figure}[H]
\centering
\vspace{1cm}
\includegraphics[width=0.5\columnwidth]{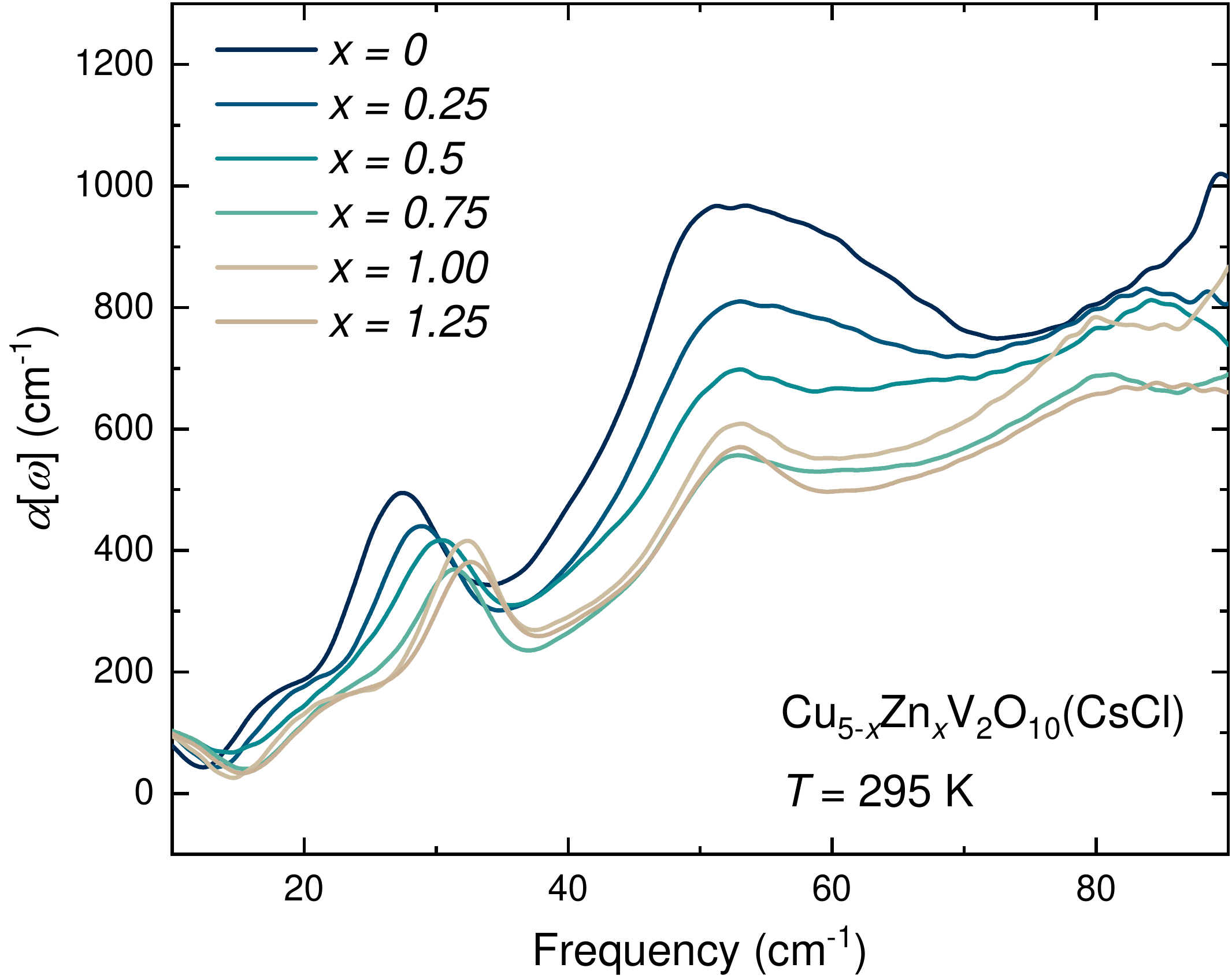}
\caption{\label{Fig_Zn} Room temperature ($T=295$ K) absorption coefficient for different Zn concentrations.}
\end{figure}

In order to investigate the influence of Zn substitution on the structural properties, room temperature measurements have been performed. Results are shown in Fig.~\ref{Fig_Zn}. Here, the phononic features are effected by substituting Cu to Zn. For instance, the 28 cm$^{-1}$ mode shows a blueshift with increasing the Zn content. The feature at around 50 cm$^{-1}$ sharpens for higher Zn concentrations as well. These changes are likely a combined effect of the smaller atomic diameter of Zn compared to Cu causing a stronger bonding force and the different space groups of the samples.

\newpage

\section{Magneto-THz measurements}
\begin{figure}[H]
\centering
\vspace{0.5cm}
\includegraphics[width=0.75\columnwidth]{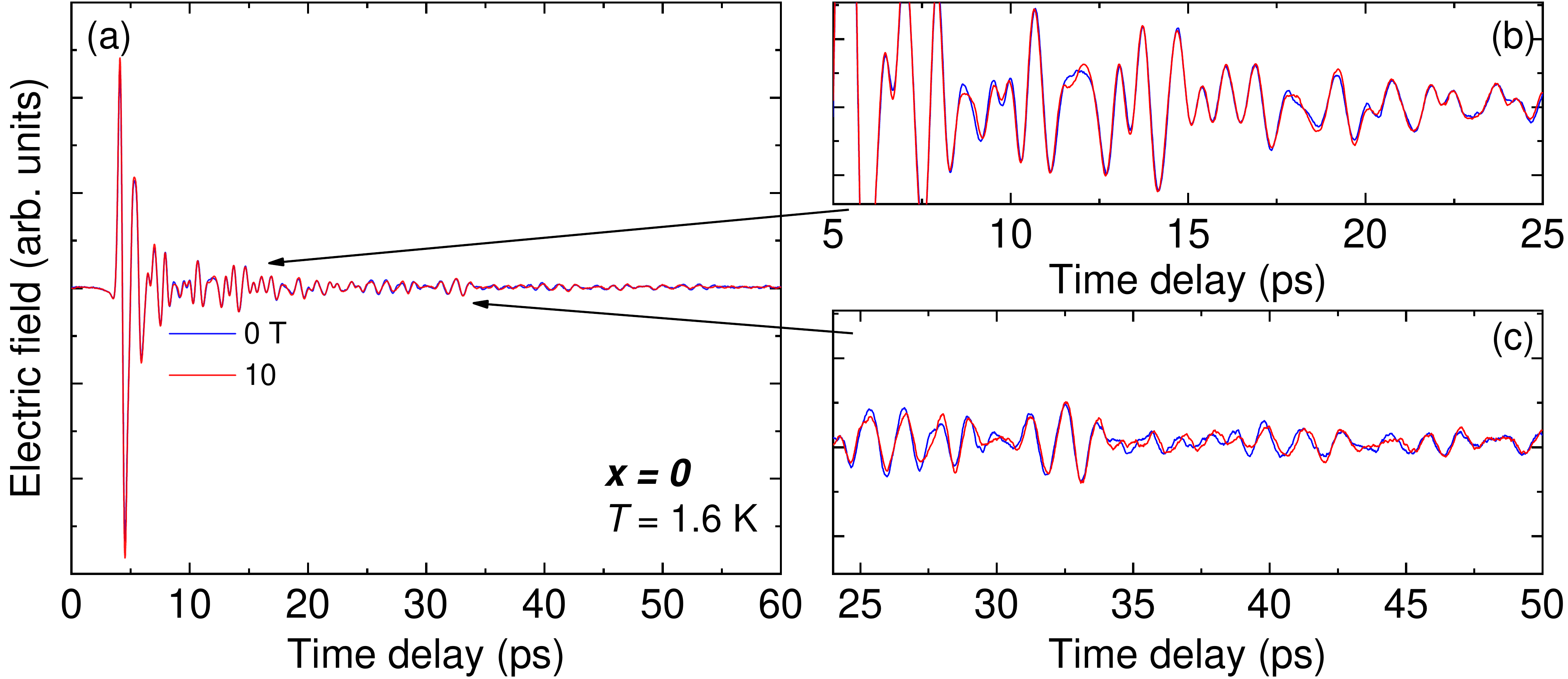}
\caption{\label{Fig_sup_magn}(a-c) Magneto-THz measurements, electric field as a function of time delay for the $x=0$ compound at $T=1.6$ K.}
\end{figure}

\begin{figure}[H]
\centering
\hspace{0.5cm}
\includegraphics[width=0.8\columnwidth]{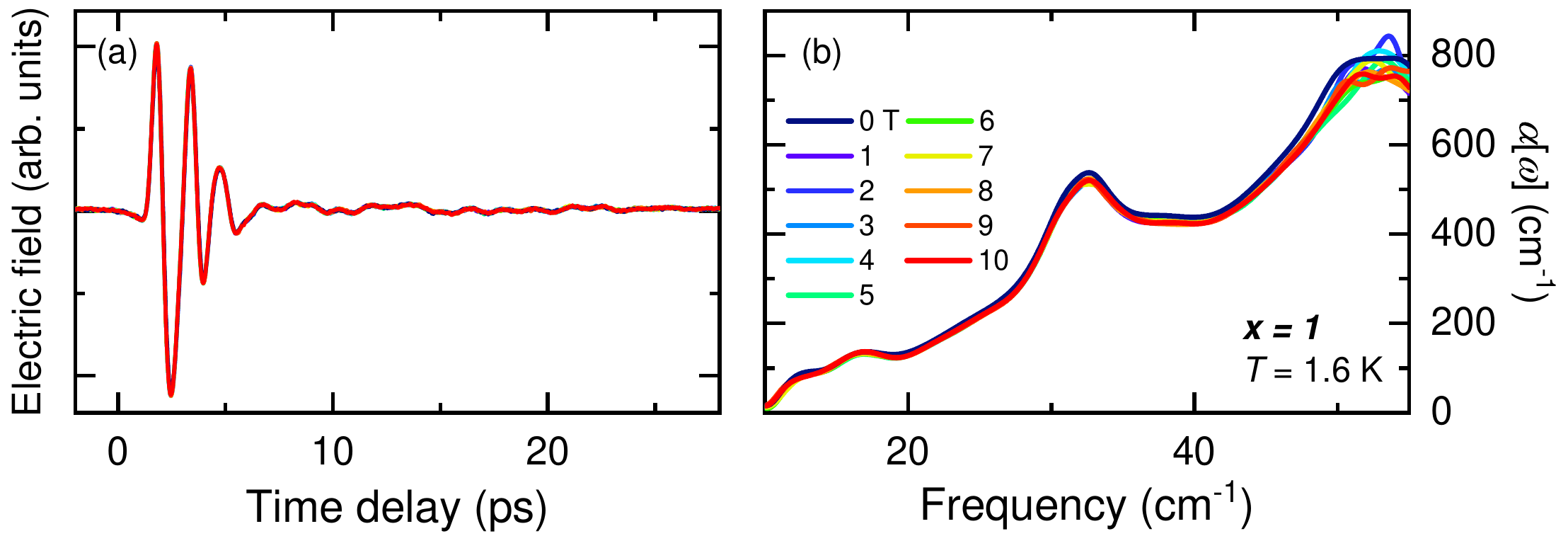}
\caption{\label{Fig_sup_x1}Magneto-THz measurements of the $x=1$ compound at $T=1.6$ K. (a) Time-domain spectra and (b) corresponding absorption coefficient, $\alpha(\omega)$. No systematic changes under magnetic fields up to $10$ T are observed.}
\end{figure}

Magneto-THz measurements were performed in Faraday geometry ($\vec{k}\parallel \vec{H}_{ext},$ $\vec{H}_{THz} \perp \vec{H}_{ext}$, where $\vec{k}$ is the wave vector of the incident THz light, $\vec{H}_{ext}$ the external magnetic field, and $\vec{H}_{THz}$ the magnetic component of the THz light) at a base temperature of $T=1.6$ K. The spectra are presented in the main text. The time-domain electric fields for the $x=0$ sample are shown in Fig.~\ref{Fig_sup_magn} up to the highest applied field of $10$ T. While the main signal below 5 ps shows only minor intensity variations [Fig.~\ref{Fig_sup_magn}~(a)], systematic changes in the extended-time oscillations are observed under magnetic field [Fig.~\ref{Fig_sup_magn}~(b,c)]. Furthermore, results for the $x=1$ sample are shown in Fig.~\ref{Fig_sup_x1} in which no changes under magnetic field are observed.


\newpage

\section{Pellet preparation and reproducibility}
All powders were thoroughly ground before pressing the pellets for the optical measurements. Although the grain size is small compared to the THz wavelength and therefore only negligible contribution due to scattering is expected in the probed frequency range, several pellets were compared to check the reproducibility of the spectroscopic features. As well to exclude any effect of the random crystallographic orientation within the powder, we repeated our THz measurements on differently prepared pellets (grinding, pellet size, and thickness).

\begin{figure}[H]
\centering
\includegraphics[width=0.45\columnwidth]{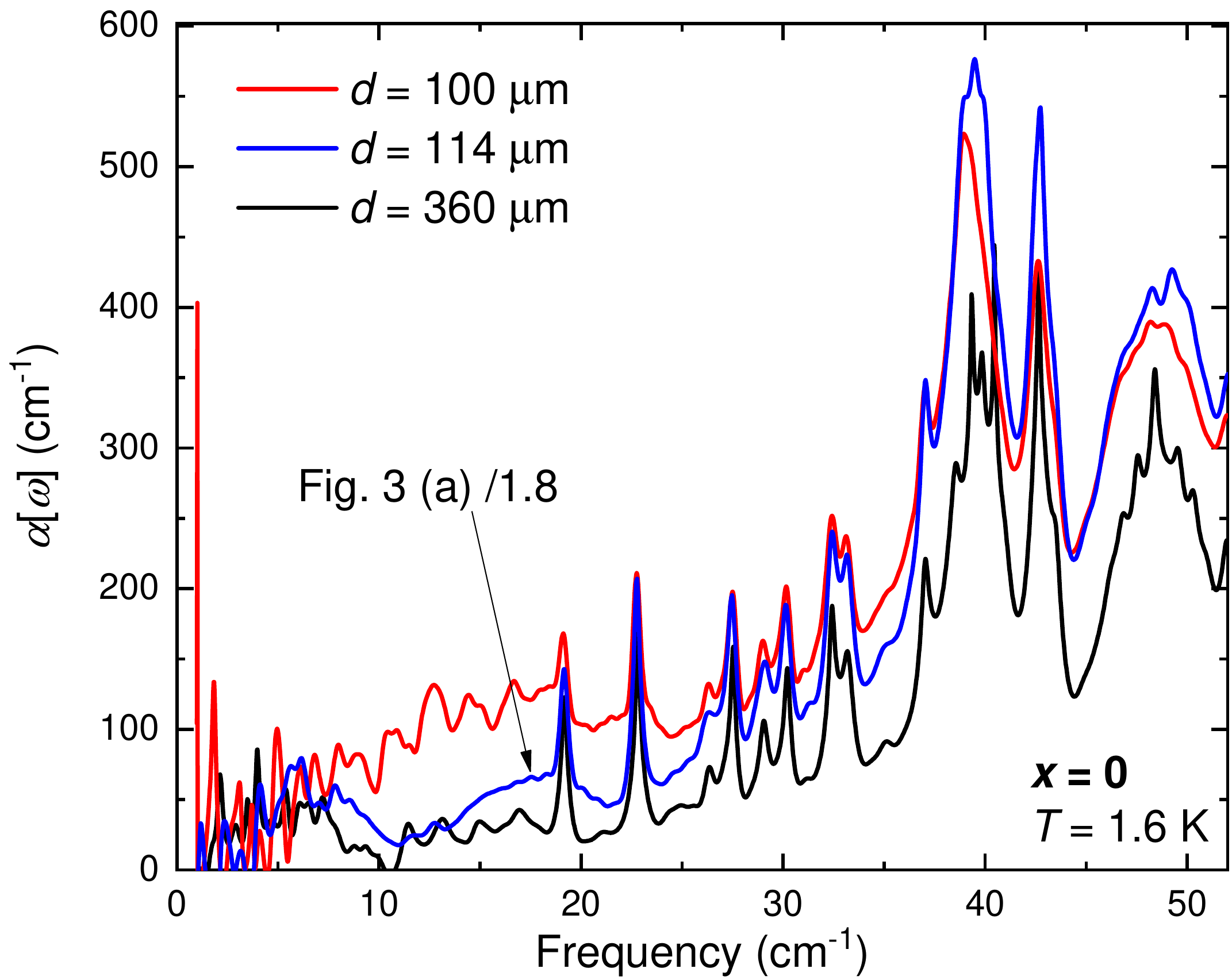}
\caption{\label{Fig_sup_comp}Absorption coefficient for different pellets. Red and black curves: 100 and 360 $\um$m thick pellets used for magneto-THz measurements. Blue curve: 114 $\um$m thick pellet from the main text. For comparison, data from the main text have been scaled.}
\end{figure}

\begin{figure}[H]
\centering
\hspace{0.8cm}
\includegraphics[width=0.6\columnwidth]{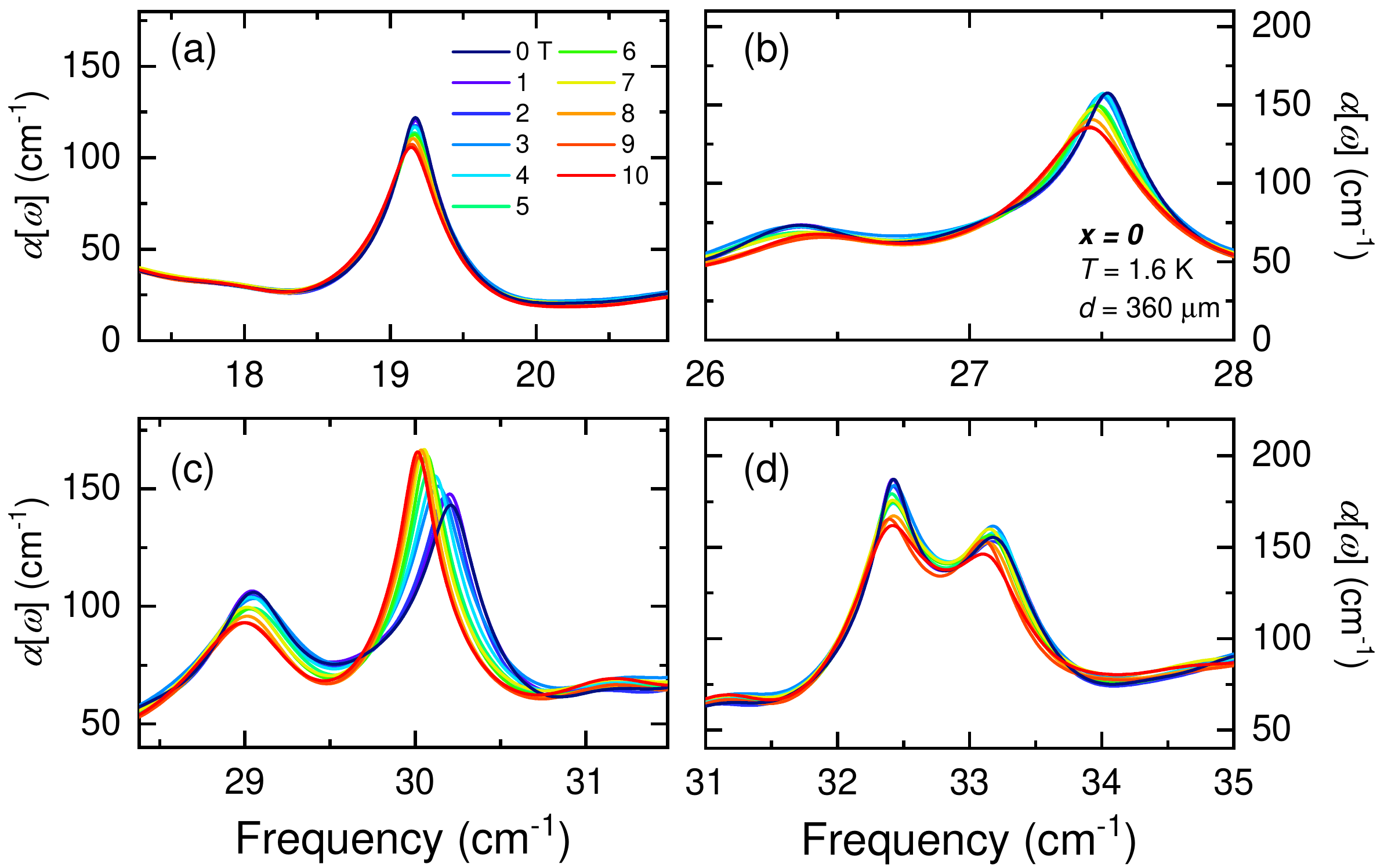}
\caption{\label{Fig_sup_thick}Magneto-optical THz TDS measurements on a 360 $\mu$m thick pellet of the $x=0$ compound at $T=1.6$ K.}
\end{figure}

In Fig.~\ref{Fig_sup_comp}, the absorption coefficient obtained from THz-TDS measurements at $1.6$ K is shown for multiple pellets. 100 and 360 $\um$m thick pellets were measured inside the superconducting magnet. To compare with the results shown in the main text (blue curve, 114 $\um$m thick pellet) the spectrum has been scaled. Overall, the main features, including the assigned resonances from the main text (15 to 50 cm$^{-1}$), show well reproducible characteristic frequencies and lineshapes. Note that the decreased signal inside the magnet compared to the measurements performed in a LiHe-bath cryostat results in slight deviations at the low energies, around 10 cm$^{-1}$. Furthermore, magneto-THz measurements were repeated for the 360 $\um$m thick pellet (see Fig.~\ref{Fig_sup_thick}). The behavior of the low-energy resonances in magnetic field is almost identical for the different pellets (cf.\ Fig.~\ref{Fig:magnetooptics}). Albeit the fact the spectra might contain mixed responses from different crystallographic orientations, other effects from the powder can be excluded. Thus, the time-domain spectra and the absorption coefficient reflect intrinsic properties of the sample.

\FloatBarrier

\end{document}